\documentclass[12pt, preprint]{aastex}

\newcommand{\be}{\begin{equation}}
\newcommand{\ee}{\end{equation}}

\newcommand{\msun}{ {\,M_\odot}}

\def\km{{\rm\,km}}

\def\sec{{\rm\,s}}

\def\gcm3{{\rm\,g\,cm^{-3}}}
\def\ncm3{{\rm\,cm^{-3}}}

\def\>{$>$}
\def\<{$<$}

\def\simlt{\lower.5ex\hbox{$\; \buildrel < \over \sim \;$}}
\def\simgt{\lower.5ex\hbox{$\; \buildrel > \over \sim \;$}}

\begin{document}

\title{Primary Versus Secondary Leptons in the EGRET SNR's} 

\medskip
\author{ Marco Fatuzzo$^1$ and Fulvio Melia$^2$} 
\bigskip 
\affil{$^1$Physics Department, Xavier University, Cincinnati, OH 45207} 
\affil{$^2$Physics Department and Steward Observatory, 
The University of Arizona, AZ 85721}
\begin{abstract} 
The EGRET supernova remnants (SNR's) are all expanding into nearby dense
molecular clouds, powering a shock at the interface where protons and
electrons accelerate to relativistic energies. A viable mechanism for
the emission of $\gamma$-rays in these sources is the decay of neutral
pions created in collisions between the relativistic hadrons and protons 
in the ambient medium. But neutral pion decay alone cannot reproduce the whole
high-energy spectrum, particularly below $100$ MeV.  A pion-decay
scenario thus requires a lepton population to ``fill in" the lower
part of the spectrum via bremsstrahlung emission.  This population, however, 
is constrained by the SNR radio spectrum.  Taking our cue from
the behavior of Sgr A East, an energetic EGRET SNR at the Galactic center,
we here examine the role played in these sources by secondary 
leptons---electrons and positrons produced in proton-proton scattering 
events and the ensuing particle cascades. We show that while secondary leptons
cannot account for the $\gamma$-rays below $100$ MeV, they can account for
the hard radio spectra observed from the EGRET SNR's.
Thus, it appears that both primary and secondary leptons may
be important contributors to the overall broadband emission from these sources,
but if so, must radiate most of their energy in 
different parts of the SNR-cloud environment.  We show that
shock acceleration in dense cores being overtaken by the expanding 
SNR shell can naturally lead to such a scenario.
\end{abstract}


\section{Introduction}          \label{sec:intro}

Several supernova remnants 
have now been associated with EGRET sources (Esposito et al. 1996;
Combi, Romero \& Benaglia 1998; Combi et al. 2001). What appears to unify 
these remnants is their interaction with nearby dense molecular clouds.  
The EGRET SNR's also constitute a subset of a larger class of 
remnants (numbering $\sim 20$) that produce OH (1720 MHZ) maser emission 
(Yusef-Zadeh et al. 2003), whose appearance helps to constrain the densities 
and temperatures of post-shock gas found in these environments.  Additionally, 
magnetic field strengths and orientations may be deduced from Zeeman splitting 
and polarization studies.  All in
all, the physical conditions within, and surrounding, these high-energy
sources, may be deduced with sufficient precision to model the diffusive
particle acceleration thought to be responsible for much of their emissivity
(as discussed in \S 2). 

By now, the EGRET SNR's have been observed in radio, infrared, X-rays and 
$\gamma$-rays. These observations provide important clues about the
processes at play in these environments.  For example, these sources
are strong radio emitters ($> 100$ Jy at 1 GHz) and have spectra that
can be significantly harder than those of most SNR's.
Additionally, the non-detection of EGRET SNR's at TeV energies 
(e.g., Buckley et al. 1998; Rowell et al. 2000) sets important constraints 
on the distribution of energetic particles. Their apparent cutoff at
high energy can be understood in terms 
of the limits placed on effective particle acceleration by the large densities
in these environments (Baring et al. 1999).  As such, the EGRET SNR's appear 
to be an ideal class of objects to study shock acceleration and cosmic ray 
production.  

Indeed, the study of gamma-ray production in SNR-cloud environments has
been undertaken by several authors.  De Jager and Mastichiadis (1997)
found that the EGRET data for W44 could result from  bremsstrahlung and 
inverse Compton emission produced by a power-law distribution of
relativistic electrons injected from a nearby pulsar
with a spectral index of $\alpha = 1.66$, 
a value chosen to also self-consistently account 
for that remnant's hard radio spectrum.  Chevalier (1999) and 
Bykov et al. (2000) fit the EGRET data of IC443 with bremsstrahlung
emission produced by nonthermal electrons injected and
accelerated in a inhomogeneous cloud environment.  Good fits to the
hard radio spectrum of IC443 ($\alpha \approx 0.36$) resulted from a
high compression ratio ($\approx 4.4$) and, to a lesser extent, from
the effects of second-order Fermi acceleration.
Several authors have considered 
the emission of gamma-rays in SNR's resulting from the creation 
of neutral pions (and their subsequent decay) in collisions between 
shock-accelerated protons and the ambient gas
(Sturner et al. 1997; Gaisser et al. 1998; Baring et al. 1999).
In these works, photons produced in the
pionic cascade account for radiation emitted above $\sim$ 100 MeV,
and a population of accelerated (primary) electrons
was invoked in order to produce bremsstrahlung  emission 
to fill in the $< 100$ MeV portion of the EGRET spectrum.  
Sturner et al. (1997) assume free-free absorption of the synchrotron
emission by an intervening hot ($8,000$ K) ISM in order to produce
a turnover at $\sim 10^7$ Hz which could then account for the seemingly
hard spectrum of IC443.  We note, however, that this warm component
of the ISM has a filling factor of only $\sim 0.2$, with most of the
ISM, by volume, being filled by a hotter, lower density medium
(McKee \& Ostriker 1977).
Baring et al. (1999) showed that, in the presence of nonlinear shock 
acceleration, primary electrons could reproduce an unusually flat
synchrotron radio spectrum that mimics that
of IC443, but found that the resulting flux was well below 
the observed radio data.  Clearly, the connection between
the $\gamma$-ray and radio emissivities in these environments
remains unsettled.

What has been lacking in previous works is a consistent 
treatment of the secondary
leptons, themselves produced via proton-proton scatterings which, in addition
to generating neutral pions, also produce charged pions. The latter decay into 
muons and thence into electrons and positrons. 
The vast array of multi-wavelength observations may now be brought 
to bear on the question of which of these populations ---primary
or secondary--- dominate the radiative emission from these sources,
requiring a more comprehensive and self-consistent treatment than has 
been attempted up until now.

An important caveat to the study of gamma-ray production in SNR's has been the 
identification of Sgr A East, a supernova-like remnant situated at the Galactic 
center, with the EGRET source 3EG J1746-2852 (Mayer-Hasselwander et al. 1998).
Like several of the EGRET SNR's, Sgr A East has also been observed at 1720 
MHz (the transition frequency of OH maser emission).  Specifically, several 
maser spots have been resolved at the structure's SE boundary with velocities 
of $\approx 50 \;\km\;\sec^{-1}$, and an additional spot has been observed 
near the Northern arm of Sgr A West (a $\sim 6$-pc mini-spiral structure of 
ionized gas orbiting about the center) at a velocity of $134 
\;\km\;\sec^{-1}$ (Yusef-Zadeh et al. 1996).  These observations indicate
the presence of shocks at the interface between the expanding Sgr A East 
shell and the surrounding dense molecular cloud.

Interestingly, the energy required to carve out the radio synchrotron structure 
within the surrounding high density cloud during the creation of Sgr A East 
appears to have been extreme ($\sim 4 \times 10^{52}$ ergs; Mezger et al. 1989). 
An analogous viewpoint was adopted by Melia et al. (1998) in their analysis of
Sgr A East's current high-energy radiative output, which also appears to be 
very high (by a factor of $50-100$) compared to that of other SNR's.  Recent 
observations made 
with the ACIS detector on board the {\it Chandra} X-ray Observatory,
however, support the hypothesis that Sgr A East is more or less a typical
single, mixed-morphology supernova remnant with an age of $\sim 10,000$ years 
(Maeda et al. 2003), casting some doubt on the interpretation of an unusually
powerful explosion made earlier by Mezger et al. (1989).   

A possible kinship between the EGRET SNR's and Sgr A East has more recently 
been explored by Fatuzzo \& Melia (2003; hereafter FM03), who adopted the
same mechanism---the decay of neutral pions created in proton-proton 
collisions---to account for the high-energy ($> 100$ MeV) emission in the 
Galactic center remnant. However, the work undertaken by FM03 invoked two 
key elements generally ignored in earlier treatments of pion-decay scenarios. 
The first is a correct treatment of the energy-dependent
pion multiplicity and its effect on the spectral shape of the pion-decay 
photons (Markoff et al. 1999; Melia et al. 1998).  The second is the 
self-consistent use of secondary leptons (by-products of charged pion 
decay) in the determination of the broadband emission spectrum. The 
importance of these secondaries becomes evident with a proper treatment 
of the pion multiplicity, as each relativistic proton produces on average 
40 - 60 leptons.  As a result, the ratio of secondary leptons to 
shock-accelerated protons can be significantly higher than has previously 
been estimated using an assumed pion-multiplicity of three. This brings 
into question earlier work on pion-production mechanisms that ignored 
the role of these decay products.  Indeed, the decay of charged pions into 
muons and, subsequently, into ``secondary" relativistic electrons and 
positrons, results in a bremsstrahlung emissivity that contributes to 
the observed spectrum below $\sim 100$ MeV.  And very importantly, 
secondary leptonic synchrotron emission also self-consistently accounts for the unusually 
steep ($\alpha \sim 1$) radio spectrum detected from Sgr A East's periphery. 

The purpose of this paper is to determine what role, if any, secondary
leptons play in the EGRET SNR's. As we shall see, 
secondary leptons alone cannot explain the broadband spectrum from these
sources. Our results suggest that {\it both} primary electrons and 
secondary leptons must play an important role. 
We discuss the SNR-molecular cloud environment in \S2.  We then
consider the role of secondary leptons produced in such 
an environment in \S3, using the EGRET SNR IC443 as a case study. 
We show that while secondary leptons can account for the unusual radio spectrum
of IC443, they alone cannot produce a pion-decay scenario
that is fully consistent with the EGRET data.
As such, we consider the role of primary electrons in IC443
in \S4, and present a viable scenario in which both primary and
secondary leptons contribute to the broadband emission of this
SNR.
In \S5, we investigate whether this scenario can also account
for the broadband observations of the EGRET SNR's W44, W28, 
and $\gamma$-Cygni.  We summarize our work and provide concluding
remarks in \S 6.

\section{The SNR-Cloud Environment}
Molecular clouds contain a total mass of $\sim 10^5 \msun$ within 
a radius of $\sim 20$ pc, thereby having average densities of $\sim
100$ cm$^{-3}$.  However, these regions of the interstellar medium
are highly nonuniform, exhibiting hierarchical structure 
that can be characterized in terms of clumps and dense cores 
surrounded by an interclump gas of density $\sim 5 - 25$ cm$^{-3}$.  
Clumps have characteristic densities of 
$\sim 10^3$ cm$^{-3}$ and radii ranging from $0.1 - 1$ pc.  In all,
these dense regions occupy a relatively small fraction (2 - 8 \%) 
of the cloud volume, but can account for most of its mass
(e.g., Williams et al. 1995).  More specifically, while a cloud may 
contain $\sim 10^3$ small ($\sim 1 \msun$) clumps, 
a significant fraction of the total cloud mass is located in a 
relatively small number ($\sim 20$) of large, massive ($\sim 10^3 \msun$)  
clumps.  These massive clumps, in turn, are comprised of
$\sim 100 - 1000$ small ($\sim 0.1$ pc), dense ($\sim 10^4 - 10^5$ cm$^{-3}$) 
cores (Lada et al. 1991; Jijina et al. 1999).  
While occupying a small total volume, these cores may contain as much as 
19\% of the cloud mass.  

The observation of OH maser emission from SNR's that are interacting
with molecular clouds has provided important clues about the
SNR-cloud environment.  High-resolution observations of W28, W44
and IC443 lead to the detection of 41, 25, and 6 individual maser
spots, respectively, in each of these three remnants
(Claussen et al. 1997).  These spots
appear to lie along, but displaced behind, the outer (leading) edge of 
the radio continuum emission.  Since OH emission is associated with
densities $10^4$ cm$^{-3} < n < 5 \times 10^5$ cm$^{-3}$,
these spots are presumably located
within dense cores that have been overtaken by the expanding SNR
shell.  

The presence of maser emission has also provided valuable
information about the magnetic fields in the SNR-cloud
environments.  
In the simplest case, where flux freezing applies,
the magnetic field strength in the interstellar 
medium scales as $B\propto \rho^{1/2}$.  
Indeed, an analysis of magnetic field strengths measured 
in molecular clouds yielded a relation between $B$ and $\rho$ 
of the form
\be
B \sim 0.1 \,\hbox{\rm mG} \left({n \over 10^4\, \hbox{\rm cm}^{-3}}
\right)^{0.47}\;,
\ee
although there is a significant amount of scatter in the data
used to produce this fit (Crutcher 1999).
This result is consistent with the
idea that nonthermal linewidths, which are measured
to be $\sim 1$ km s$^{-1}$ (e.g., Lada et al. 1991), 
arise from MHD fluctuations.
The observed  OH (1720 MHz) Zeeman splitting in the EGRET
SNR's yields line of sight magnetic field strengths that range
from $0.1  - 4$ mG (Claussen et al. 1997;  Koralesky et al. 1998;
Brogan et al. 2000).  While such field strengths can arise in the
dense regions associated with the OH masers, these measurements
may also point to an enhancement of the magnetic field, perhaps arising from
shock compression.  Indeed,
Chevalier (1999) determined that the magnetic field
strength in the SNR's radiative shell after compression is
\be
B = 0.24\, n_1^{1/2}\, v_2\,\, \hbox{\rm mG}\;,
\ee
where $n_1$ is the interclump density in units of
10 cm$^{-3}$ and $v_2$ is the shock velocity in units
of 100 km s$^{-1}$.  Interestingly, Claussen et al. (1997)
derived line-of-sight magnetic fields of order 0.2 mG over
regions that were several parsecs apart in both
W28 and W44. It should be noted, however, that 
the OH measured field strength may
overestimate the fields by as much as a factor of 5
(Brogan et al. 2000).

Calculating the age of SNR's expanding into a nonuniform medium
is difficult, as can be witnessed by
the large discrepancy in the ages determined for specific
remnants.  For example, estimates for the age of 
W44 range from 6000 - 7500 year (Rho et al. 1994)
to $\sim 29,000$ years (Koo \& Heiles 1995).
Likewise, the age of IC443 has been estimated to be
$\sim 1,000$ years (Wang et al. 1992), 3,000 year 
(Petre et al. 1988), and 30,000 years (Chevalier 1999).
As elaborated upon below, a scenario whereby secondary leptons  
contribute significantly to the broadband
emission of the EGRET SNR's favors an age of 30,000 years.
We therefore adopt this age throughout the rest of the paper.

\section{The Role of Secondary Leptons in IC443}
We demonstrate in this section that while a pion-decay 
mechanism with only secondary leptons can account 
for the high-energy spectrum of Sgr A East, a similar mechanism apparently
cannot do so for the EGRET SNR's.  More specifically, while the high-energy 
($>100 $ MeV) portion of the latter's spectrum can be fit by the decay of 
neutral pions produced by collisions between power-law, shock-accelerated
protons and the ambient medium, secondary leptons produced concomitantly
via the decay of charged pions cannot reproduce the observed $<100$ MeV 
emission.  However, secondary leptons can account for the unusually 
hard radio spectra of the EGRET SNR's.

There are two aspects of the secondary lepton distribution that make it
incapable of accounting for the $< 100$ MeV emission in the EGRET SNR's.
First, the maximum power radiated by leptons is not expected 
to greatly exceed the luminosity generated via the decay of neutral 
pions, since
the charged pion production rate mirrors (within a factor two) that of
the neutrals.  Second, leptons are injected into the system with a 
distribution that peaks near $50$ MeV (see Figure 2 of FM03). Since  
cooling via electronic excitation dominates over bremsstrahlung 
for energies below $\sim 100$ 
MeV, a cooled distribution of secondaries will itself have a break near 
$\sim 100$ MeV.  In turn, the high-energy emission of secondaries, which 
in the SNR environment is dominated by bremsstrahlung, steepens near $100$ 
MeV.  All these factors together preclude any possibility of the 
secondary leptons alone accounting for the whole EGRET SNR spectrum.

To clearly illustrate this point, we apply the model developed for
Sgr A East to the $\gamma$-ray data pertaining to 
SNR IC443.  Since the model is discussed extensively in FM03, we will 
not reproduce the details here. However, we note that for this
study, we assume that the medium is comprised of neutral H, and hence adopt
the appropriate bremsstrahlung and electronic excitation cooling rates
given in Gould (1975).  In addition, we use the atomic bremsstrahlung
cross-section of Blumenthal \& Gould (1970) to calculate the
bremsstrahlung emission.  We note, however, that the choice of a neutral
medium over a fully ionized one has little effect on our results.

The high-energy portion of the spectrum 
is fit quite well with the pion bump resulting from  
shock-accelerated protons characterized by a power-law
distribution $dn_p/dE \propto E^{-\alpha_p}$ with a spectral index $\alpha_p = 
2.2$ (that falls within the expected range of 2 - 2.4).  
However, the non-detection of the EGRET SNR's by Whipple (Buckley et al.
1998) and
CANGAROO (Rowell et al. 2000) mandates a maximum energy $E_{max}$ for this
distribution. 
The spectrum is normalized by the product $n_H\cdot n_p 
\cdot V$, chosen on the basis of an assumed source distance 
of $1.5$ kpc. Here, $n_H$ is the proton number density in
the ambient medium, $n_p$ is the number density of shock-accelerated 
protons, and $V$ is the volume of the emitting region.

For any given SNR, the number of secondary leptons depends on 
the injection rate, the rate at which these particles either
lose energy or leave the system, and the duration over which
particles have been injected in the system. Their $\gamma$-ray emissivity, 
primarily bremsstrahlung, peaks when they reach a steady state.
The maximum bremsstrahlung emissivity below $100$ MeV is correlated 
directly with the radiative flux produced by decaying neutral pions
because both are ultimately produced by the same proton-proton 
scattering events. (Note, however, that secondary leptons that 
produce higher energy emission can cool more efficiently via 
synchrotron emission, in which case the above claim cannot be made.)

This result holds regardless of the number density $n_H$ of the ambient
medium, because the density of steady-state secondary leptons that emit 
via bremsstrahlung scales inversely with $n_H$. This 
dependence offsets the $\propto n_H$ variation of the emissivity. Of
course, the bremsstrahlung spectrum also depends upon the choice of 
magnetic field strength, but only at sufficiently high-energies 
(typically $> 10^3$ MeV) so that synchrotron cooling dominates over 
bremsstrahlung.  The resulting spectral shape below 
$10^3$ MeV is therefore completely specified by the single 
parameter $\alpha_p$, and is fit to the data through the single choice of 
the product $n_H\cdot n_p\cdot V$.  This ``universal" curve is shown 
by the solid line in Figure 1 for $\alpha_p = 2.2$ (and $E_{max}$ 
= $10^6$ MeV), together with the EGRET data and Whipple upper-limit
pertaining to SNR IC443. 
Clearly the bremsstrahlung emission due to secondary leptons alone
cannot account for the spectrum below 100 MeV---a result that also holds 
(and is even more pronounced) for  W28, W44, and $\gamma$-Cygni.  

Compounding the secondary lepton's inability to produce the
bremsstrahlung emission required to ``fill in" the sub-100 MeV
portion of the EGRET spectrum is the fact that a steady-state
assumption is difficult to justify in the SNR environment.
This point is illustrated by Figure 2, which shows the lepton cooling
time (defined as $\tau_{cool} = E / \dot E$) as a function of lepton
energy associated with bremsstrahlung (short dashed),
synchrotron (long dashed), inverse Compton scattering
with the Cosmic Microwave Background (dotted) and electronic excitation (solid)
for a neutral medium with a density
of $n = 300$ cm$^{-3}$ and field strength of $B = 0.3$ mG. 
It is clear that
even for the assumed remnant age of 30,000 years, the steady-state
assumption would require densities in excess of $\sim 1000$ cm$^{-3}$.
Such densities are well above the average cloud density and the
expected density of the radiative shell ($\sim 500$ cm$^{-3}$; see
Chevalier 1999).  It does not appear, therefore, that leptons 
have sufficient time to reach steady state.  
To illustrate this point further, we plot in Figure 3 the steady-state 
lepton distributions arising from the production of the
pion-bump shown in Figure 1 for two value of $n_H$:
300 cm$^{-3}$ and 
3,000 cm$^{-3}$.  Since the lepton injection rate is fixed
by the normalization of the pion bump to the EGRET data
(and therefore independent of the ambient density),
the steady-state distribution for the $n_H = 300$ cm$^{-3}$
case (dotted line) attains a higher value than its high-density counterpart
(dashed line).  The solid line represents the lepton distribution 
attained by simply multiplying the injection rate by an age of 30,000
years.  As can be seen, the built-up injected distribution 
falls below the lower-density
steady-state distribution over most of the energy range.
As such, a more realistic secondary lepton emissivity
is represented by the dotted line in Figure 1,
which shows the bremsstrahlung emission produced by the secondary leptons
associated with the lower-density case.  For simplicity, we have
assumed that the true particle distribution for the lower-density
(non steady-state) case is given by the lower of the solid and dashed
curves in Figure 3, and adopt such a convention throughout this 
paper.

An underlying assumption made in the above discussion of the 
lepton distribution is that the injection rate is constant in
time.  In turn, this assumes that the proton distribution and
the ambient medium with which it interacts do not change over the
remnant's evolution. 
One could speculate an earlier brief period of significantly
higher injection, so that the lepton distribution assumed
above significantly underestimates the true lepton population.
However, such an epoch would imply that a considerably larger 
number of relativistic protons permeated the region
during this highly active era ($\sim 10^2$ greater than
is presently inferred).  Such a scenario is therefore
energetically unfavorable.  Specifically, for a distance
of 1.5 kpc to IC443, the energy content in relativistic protons
that produce the pion-bump shown in Figure 1 is 
\be
E \approx 0.004\times 10^{51}\,\hbox{ergs}\,\left({n_H\over
300\,\hbox{cm}^{-3}}\right)^{-1}\,.
\ee
A significantly larger proton population would thus seem 
energetically untenable.  In addition, one would also need to
argue that protons have diffused out of the SNR environment, whereas
the leptons did not.

While secondary leptons do not appear capable of
accounting for the observed gamma-ray emissivity, they
do appear to be capable of accounting for the unusually
hard radio spectra observed from the EGRET SNR's.   
To explore this issue, we consider the synchrotron
emissivity of secondary leptons produced under
a baseline set of parameters
$\alpha_p = 2$, $E_{max} = 10^6$ MeV, $B = 0.29$ mG, and
$n_H = 300$ cm$^{-3}$ (case A).
We sample the parameter space around our baseline point by
considering the following additional cases:  B) same as A
but with $E_{max} = 10^5$ MeV; C) same as A but with 
$\alpha_p = 2.4$; D) same as A but with $n_H = 3000$ cm$^{-3}$ and
$B = 0.45$ mG.  All cases have an assumed age of 30,000 year, and
are normalized to the EGRET data as shown in Figure 4.  
The resulting synchrotron emissivity
for these cases is then shown in Figure 5, where
the radio data for IC443 were taken from  Erickson \&
Mahoney (1985).  Solid circles refer to data that were 
corrected via a calibration factor, whereas open circles
refer to data for which such a correction was not possible.
The effect of free-free
absorption by the warm component of the ISM is not included in the 
radio fits.  Since this component has an expected electron density 
of $n_{ew} = 0.17$ cm$^{-3}$ and a temperature of $T = 8000$ K
(McKee \& Ostriker 1975), 
the resulting absorption coefficient is $\alpha_{ff}
\approx 6\times 10^{-23}$ at a frequency of $\nu = 10^7$ Hz
(see, e.g., Sturner et al. 1997).
For a source distance of 1.5 kpc and a filling factor of $\sim 0.2$,
free-free absorption is unlikely to
have a significant effect on the radio spectra of IC443 above $\sim 10^7$ Hz, 
although it may be responsible for the apparent low-energy turnover
at around that frequency as suggested by the data presented in Figure 5
(see also Figure 14).

There are several important constraints that the radio
data place on the pion-decay scenario.  Specifically, 
if the radio emission does indeed arise from secondary
leptons, then such an association would favor a proton
distribution with a spectral index of 2 and a high-energy
cut-off of $\sim 10^6$ MeV.  In addition, this association
clearly favors a lower-density ($< 10^3$ cm$^{-3}$) 
environment and a remnant age of $\sim 30,000$ year.   
Although a younger age can be
adopted, doing so would lower the number of secondary
leptons, and hence, would require a stronger magnetic field
to compensate.  Adopting a younger age would then be in
conflict with the measured fields in the EGRET SNR's of
$\sim 0.2$ mG (see \S 2).  We note that for densities
below $10^3$ cm$^{-3}$, the lepton distribution over
the energy range responsible for the observed radio emission
is independent of the ambient density since it is just
the injected distribution built up over the age of the remnant.
Once one reaches sufficiently high densities, the lepton
distribution is then set by its steady-state value, and
scales as the inverse of $n_H$ (see Figure 3).  As a result, the
magnetic field required for case D is greater than those
of cases A - C.
 
It is quite intriguing that the radio data are fit so well by case
A since such densities and field strengths are expected
in the radiative shells of SNR's interacting with molecular
clouds (Chevalier 1999).  In addition, a 0.29 mG magnetic
field is consistent with observations (see \S 2 above).

We conclude from the analysis presented in this section
that secondary leptons produced in a pion-decay
scenario cannot account for the sub-100 MeV $\gamma$-ray emission
required to make such a scenario compatible with the EGRET
data over all energies.  However, this population of leptons can
seemingly account for the unusually hard radio spectra observed from the
EGRET SNR's.  Although the focus of our discussion thus far
has been on IC443, our conclusions also hold for
W28, W44 and $\gamma$-Cygni (as further discussed in \S5).

\section{The Role of Primary Electrons in the Nonhomogeneous Medium
of IC443}
The fact that secondary leptons cannot reproduce the $\gamma$-ray spectrum
of the EGRET SNR's below $100$ MeV suggests the presence of an additional 
electron population in these sources.  Yet the radio data are explained rather
well by the former, so the physical conditions where these other particles 
operate are severely constrained in terms of their synchrotron emissivity.
At best, a combination of primary and secondary leptons may be responsible
for the radio emission.

Let us adopt the view that the diffusive shock mechanism responsible for 
accelerating the protons also energizes a power-law distribution of 
(``primary") electrons.  From previous works (e.g., Gaisser
et al. 1998), it is known that the EGRET
data from IC443 can be reproduced by a scenario in which both
the primary electrons
and protons have a spectral index of $\alpha = 2.4$.  However,
the resulting secondary leptons would then be too steep to provide a
good fit to the radio data (see Case C in Figure 5).  
Furthermore, it is not difficult to see that the $\gamma$-ray spectra
of the EGRET SNR's are
actually compliant to an even steeper primary electron
distribution.  We therefore consider the possibility that
the primary electron population is injected with a power-law distribution
$\propto E^{-\alpha_e}$ whose spectral index $\alpha_e$
is different from $\alpha_p$.

In Figure 6 we show fits to the $\gamma$-ray spectrum of IC443
for our pion-decay scenario with a proton spectral index of
$\alpha_p = 2$ and cutoff energy of $10^6$ MeV, and an ambient
medium of density $n_H = 300$ cm$^{-3}$ and field strength
$B = 0.29$ mG (case A above), but with the addition of
bremsstrahlung emission from
a power-law distribution of primary electrons in equipartition
with the proton distribution, and normalized to the sub-100 MeV
EGRET data.  The latter constraints lead to a value of $\alpha_e = 2.8$.  
Such a steep electron distribution almost
certainly implies nonlinearity in the shock.  While the primary
electron distribution index $\alpha_e$ is not well constrained, 
it is not clear whether
$\alpha_p$ and $\alpha_e$ can differ by so much. 
Still, values of $\alpha_e$ this 
large have emerged from both analytical and (highly detailed) numerical 
simulations of particle acceleration in certain energy ranges, so they 
cannot be ruled out easily (see, e.g., Baring et al. 1999; Ellison et 
al. 2000). 

While the assumed population of primary leptons can provide the necessary
bremsstrahlung emission to produce a good fit to the EGRET data, as shown
in Figure 6, 
it is not compatible with the radio data if located in the
same region as the secondary leptons.  This point is illustrated in Figure
7, which shows the synchrotron emissivities of the secondary
leptons (dashed curve) and the primary electrons (solid curve) 
for a common field strength of 0.29 mG.  This result suggests that
the inhomogeneity of the SNR-cloud environment results in two distinct
emission zones for the primary and secondary populations.

Taking our cue from the observation of OH masers (see \S 2),
we consider the possibility that particle acceleration (both
proton and electron) occurs as dense cores are overtaken 
by the expanding radiative shell.  Interestingly, it takes
$\sim 10^3$ years for a shell moving at $\sim
100$ km s$^{-1}$ to sweep past a 0.1 pc core.  With the radiative shell
expanding several parsecs into the cloud, an appreciable fraction of 
the roughly $10^4$ cores would have been overtaken, and thus,
have been active acceleration sites.  An active epoch
lasting $\sim 10^2 - 10^3$ years in a given core could thus be consistent with
the number of observed maser spots in W44, W28, and IC443
(see \S 2).  Indeed, these observations would suggest that
roughly 500 to 5,000 cores have been activated in each of these regions.

In these high-density environments ($\sim 5\times 10^4$ cm$^{-3}$), 
the accelerated electrons would cool in $\sim 300$ years.  If the time
for particles to diffuse out of these dense regions was appreciably
longer than the cooling time (but shorter than the age of the remnant),
then the electrons would efficiently radiate most of their
power while still in these regions.
On the other hand, the protons in these cores have a cooling time 
of order $\sim 10^3$ years and could
therefore diffuse into the lower density radiative shell
before losing an appreciable amount of energy.  
In order to explore this scenario, we consider the injection
of electrons and protons into cores with density of $5\times 10^4$ cm$^{-3}$
and field strength B = 0.22 mG (from Eq. 1).  We assume that
the protons diffuse into the lower density radiative shell, thus
reproducing the scenario explored in case A above.  In contrast, the
electrons are assumed to radiate most of their power in the
dense regions before diffusing out.  We approximate the resulting
bremsstrahlung emission by calculating the steady-state distribution
produced by the injection of power-law electrons with
$\alpha_e = 2.8$ in equipartition with the proton distribution, 
but confined to the high-density cores.
The resulting $\gamma$-ray emissivities are presented in
Figure 8, with the short and long dashed curves as in Figure 6,
and the dotted curve representing the ensuing primary electron
bremsstrahlung emissivity.   
Because electronic excitation effectively cools the electrons below 100 MeV, 
the steady-state distribution is less steep in the lower-energy
range and, as a result, produces a bremsstrahlung emissivity
that does not fit the lowest energy data point.  

Given the nature of nonlinear shocks, it may be 
possible that the electron distribution steepens significantly
at energies $>$ 10 MeV, thereby providing a good fit to the
EGRET data without violating the imposed equipartition
constraint.  Indeed, the resulting $\gamma$-ray 
emissivity (calculated in the same manner as for Figure 8)
for an electron distribution with $\alpha_e = 3.4$
is shown in Figure 9, demonstrating the viability of this
scenario in accounting for the observed spectrum of IC443.
The radio spectrum of the primary electrons is approximated 
by assuming that the steady-state distribution derived in the
high-density medium also describes the electrons after they 
have diffused into the radiative shell.  The resulting
synchrotron emissivity for the primary electrons
(dashed curve), along with
the emissivity of the secondary leptons (solid curve), is shown in Figure 
10.  Clearly, the primary electrons cool effectively in the 
high-density environment before diffusing out into the shell, 
so that their synchrotron emissivity 
falls well below that of the secondaries.

The total energy content of the proton distribution
required to fit the EGRET data is given by Equation 3,
and therefore represents a small fraction of the overall
blast energy.
However, the acceleration of particles at numerous
localized sites places stronger constraints on the
model energetics.  For example, with an assumed
activation of $\sim 5000$ cores, each with radius $\sim 0.1$ pc,
the total cross-sectional area of ``active sites" energized
by the expanding shell is roughly 0.06\% of the total shell's surface area
(for an assumed shell radius of 15 pc).
Clearly, this scenario thus requires a very high particle acceleration
efficiency at the shell-core interface.  

\section{The Two-Zone Emission Model: Application to W28, W44 
and $\gamma$-Cygni}
In the previous section, we presented
a two-zone emission model to self-consistently account
for both the $\gamma$-ray and radio data observed from IC443.
In this section, we apply this model to W28, W44, and $\gamma$-Cygni.
Although there are several free parameters that can be varied, 
we set the proton distribution high-energy cut-off, the
primary electron spectral index, 
the shell density, the core density, and the remnant age to
values of $E_{max} = 10^6$ MeV, $\alpha_e = 3.4$,
$n_H = 300$ cm$^{-3}$,
$n_{Hc} = 5\times 10^4$ cm$^{-3}$, and 30,000 year, respectively.
We treat $\alpha_p$ and $B$ as free parameters, and normalize
the proton and primary electron distributions through fits to the
EGRET data.

The results for W44 are shown in Figures 11 - 12, where the
free parameters are $\alpha_p = 2.0$ and  $B = .43$ mG.
The results for W28 are shown in Figures 13 - 14, where
the free parameters are identical to those chosen for 
W44.  The results of $\gamma$-Cygni are shown in Figures
15 - 16, where $\alpha_p = 2.2$ and $B = .28$ mG.
In all, the two-zone model presented here provides good fits
to the radio and $\gamma$-ray data observed from the
EGRET SNR's.  

\section{Conclusions}
An earlier analysis of the high-energy emission from Sgr A East has shown
that its radio and $\gamma$-ray spectra may be understood self-consistently 
in the context of neutral pion production in proton-proton scatterings and 
their subsequent decay into photons and relativistic (secondary) leptons.  
The latter, in particular, account for this remnant's radio emission
with the correct spectral shape and flux, once the ambient physical 
conditions are understood from fits to the $\gamma$-ray data.

A similar analysis of the other EGRET SNR's has now revealed that 
secondary leptons may account for most (if not all) of the
SNR radio spectrum. Interestingly, the environment required to
produce good fits to the radio data are consistent with those
expected from a dense ($n_H\sim 500$ cm$^{-3}$) and
highly-magnetized ($B\sim 0.3$ mG) shell, and is consistent with
the observed magnetic field strengths inferred from these regions.
However, a pion decay scenario with only secondary leptons cannot account 
for the observed $\gamma$-ray spectra.
It therefore appears that a population of shock-accelerated
electrons must also be present.

Given the constraints imposed on the primary electrons by the radio
data, and taking our cue from the observation of several OH maser
spots observed in the SNR environments, we have presented a two-zone model
wherein protons and electrons are accelerated in the dense cores
of molecular clouds that are energized by the expanding SNR shell.
While as many as $\sim 5,000$ such cores may have been energized 
in these environments, a relatively short ``active" era 
(lasting $\sim 10^2 - 10^3$ years) could then explain the observed
number of OH maser spots.  In addition, if the electron cooling time
in these dense regions ($\sim 300$ years) is shorter than the
diffusion time, these electrons would cool effectively within the
cores before diffusing out into the radiative shell.  Their resulting
synchrotron emissivity would then fall below the observation limits.
In contrast, the longer cooling time for the protons could allow them
to diffuse out into the shell before producing pions via $p-p$ collisions
with the ambient medium.  

We find that the two-zone model provides good fits to the radio and
$\gamma$-ray data for IC443, W44, W28 and $\gamma$-Cygni so long as
the primary electrons are injected with a steep $\alpha_e = 3.4$
power-law distribution above 10 MeV (in order to not violate the
imposed condition that this distribution be injected with the
same energy as the proton distribution.)  Such a distribution would
then signal the importance of nonlinear effects in shock acceleration.
In addition, our model suggests that the EGRET SNR's 
are $\sim 30,000$ years old.
While at the high end of the inferred ages for these remnants, this 
result may help explain why most of the SNR's that also produce OH
maser emission (and are therefore interacting with dense regions
of molecular clouds) have not been observed in $\gamma$-rays,
as they may not be old enough to have built up populations of
relativistic particles capable of producing an observable emissivity.
Finally, while the total energy in relativistic particles 
represents a small ($<$ 1\%) fraction of the total blast energy,
the association of acceleration sites with dense cores suggests
that a large percentage of the local energy
($\sim 10$\%) must be converted into nonthermal particle energy.

An outstanding issue raised by this work is whether the two-zone
model invoked to account for the broadband emission of the EGRET
SNR's is consistent with the picture developed by FM03 
for Sgr A East.  In the work of FM03, particle diffusion out of the
Sgr A East environment was required to produce the steep lepton
distribution inferred from the radio data.  Such a diffusion process
was not required in this work.  However, recent observations of
the Galactic center by HESS (Aharonian et al. 2004) 
suggest that there may be two 
acceleration regions in Sgr A East
(Crocker et al. 2004), thereby complicating the overall
picture.  In addition, the possible
association of the AGASA and SUGAR anisotropies 
at energies $\sim 10^{18}$ eV near the Galactic center
with particle acceleration at Sgr A East (Crocker et al.
2004) casts some doubt
on the diffusion scenario invoked by FM03.
In light of this work and recent observations, 
a detailed re-examination of the pion-decay model in Sgr A East 
is therefore warranted. 

\bigskip 
\centerline{\bf Acknowledgments} 
We are grateful to Matthew Baring and Fred Adams for very helpful discussions.
This research was supported in part by NASA grant NAG5-9205 and NSF 
grant AST-0402502 at the University of Arizona. MF is supported by the 
Hauck Foundation through Xavier University.

\clearpage
\newpage
\begin{figure}
\figurenum{1}
{\epsscale{0.9} \plotone{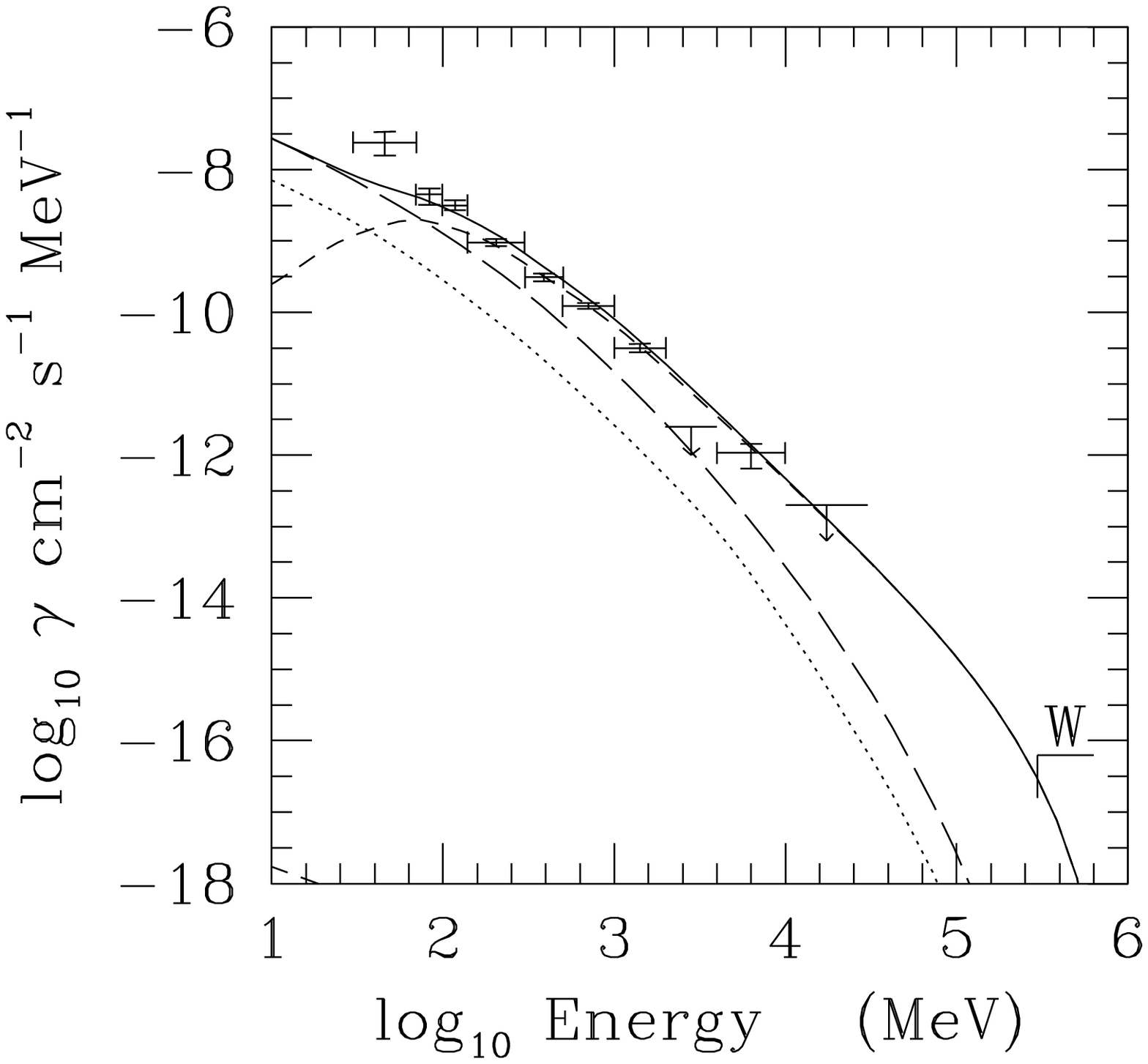} } 
\figcaption{Secondary lepton model for IC443 for a proton distribution
with a spectral index of $\alpha_p = 2.2$ and a high-energy cutoff
of $E_{max} = 10^6$ MeV . Short dashed curve: 
photons produced via neutral pion decay from proton-proton 
scattering events; long dashed curve: bremsstrahlung emission from
the secondary leptons produced via charged pion decays for
a steady-state distribution; solid curve: 
total spectrum for the steady state scenario;
dotted curve:  bremsstrahlung emission from the secondary
leptons for an ambient density of $n_H = 300$ cm$^{-3}$ and
and injection duration of 30,000 years.
Data are taken from Esposito et al. (1996) 
and Buckley et al. 1998.}
\end{figure}

\clearpage\newpage
\begin{figure}
\figurenum{2}
{\epsscale{0.9} \plotone{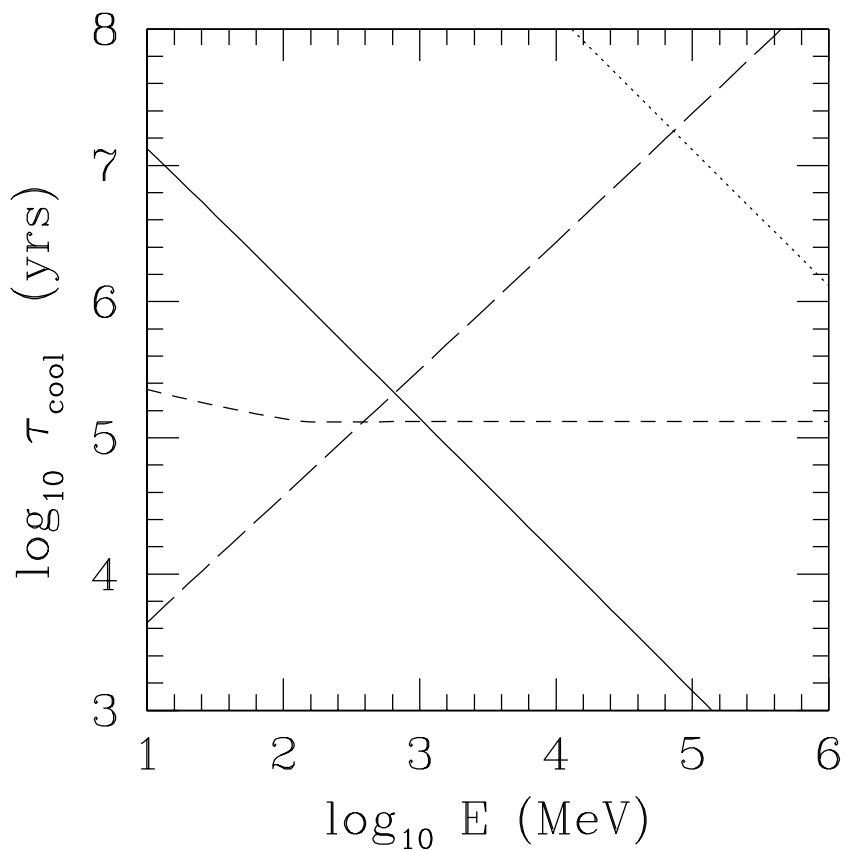} } 
\figcaption{The cooling time $\tau_{cool} =
E / \dot E$ as a function of energy 
for leptons interacting with a neutral medium of density
$n_H = 300$ cm$^{-3}$ and magnetic field strength $B = 0.3$ mG.
Short dashed curve: bremsstrahlung cooling; long dashed curve:
synchrotron cooling; solid curve: electronic excitation losses; 
dotted curve: inverse Compton scattering with the Cosmic Microwave Background.
}
\end{figure}

\clearpage\newpage
\begin{figure}
\figurenum{3}
{\epsscale{0.9} \plotone{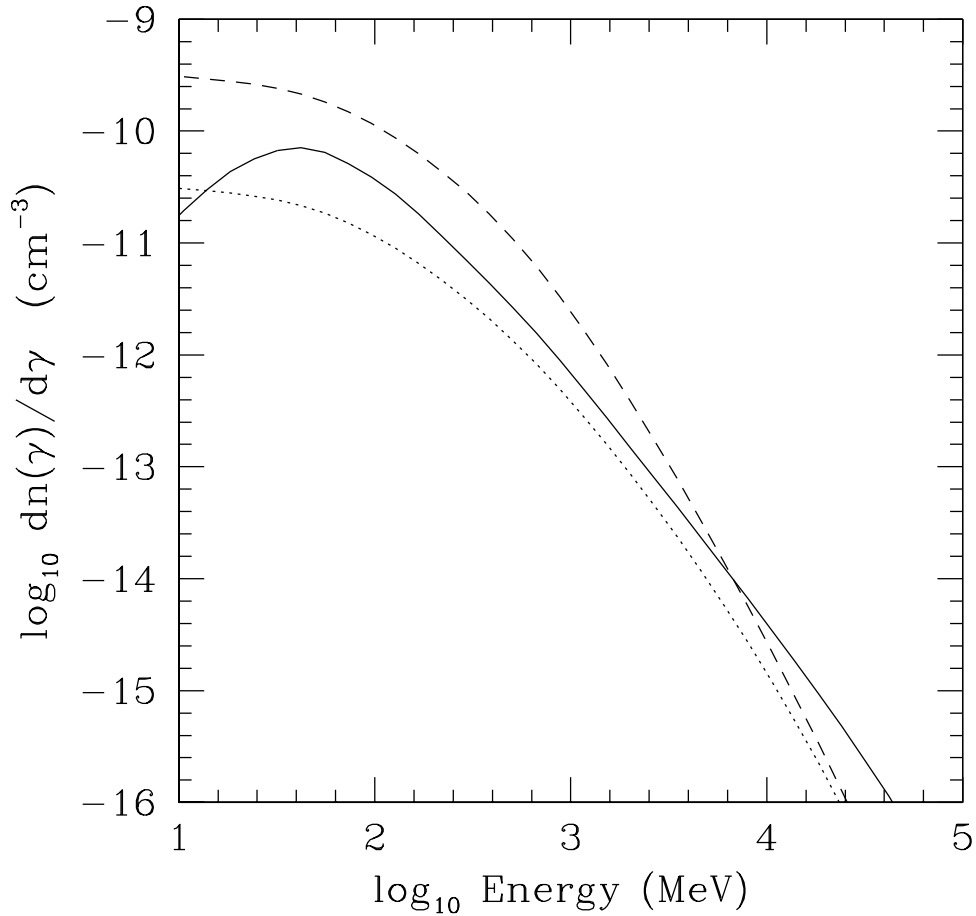} } 
\figcaption{Steady-state lepton distributions arising from the
pion-decay mechanism used to fit the pion-bump shown in Figure 1.
The dotted line depicts the ensuing steady-state distribution for
an assumed ambient density of $n_H = 3,000$ cm$^{-3}$, whereas the
dashed curve shows the steady-state distribution for a density
of $n_H = 300$ cm$^{-3}$.  The solid line represents the
lepton injection rate (which is tied to the EGRET data, and
therefore independent of $n_H$) multiplied by the assumed age
of 30,000 years.
}
\end{figure}

\clearpage\newpage
\begin{figure}
\figurenum{4}
{\epsscale{0.9} \plotone{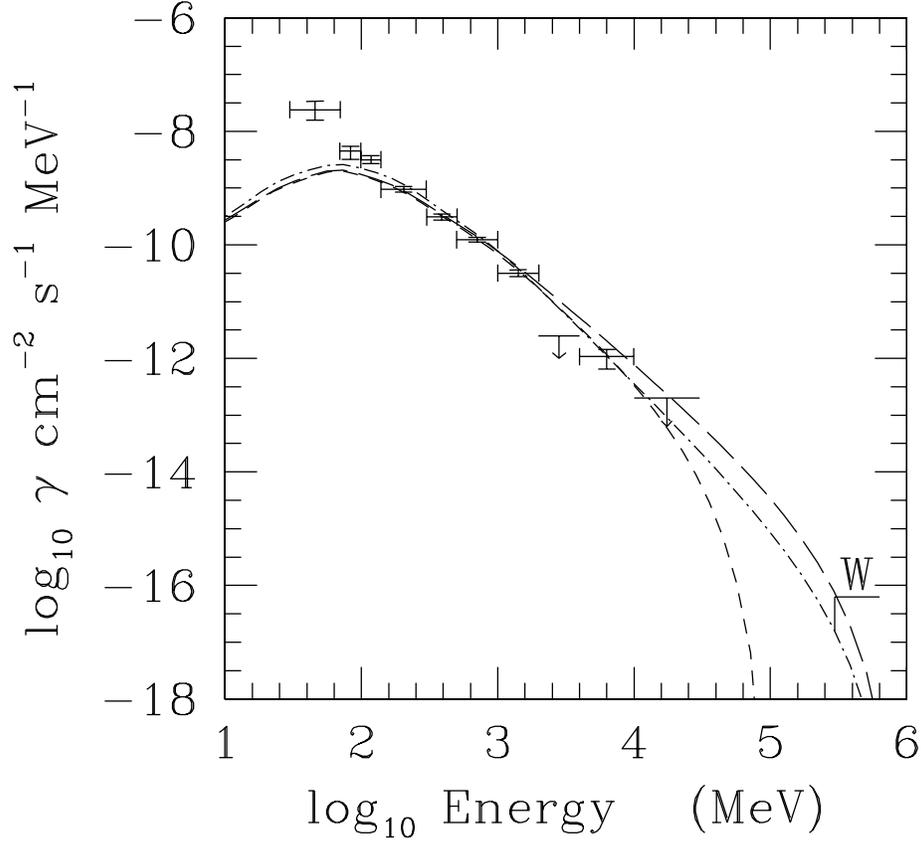} } 
\figcaption{Photons produced via neutral pion decay arising from the
interaction between a relativistic power-law distribution of protons
with spectral index $\alpha_p$ and high-energy cutoff $E_{max}$,
and an ambient medium of density $n_H$ and field strength $B$.
Fits are normalized to IC443.  Long dashed curve -- case A:  
$\alpha_p = 2.0$, $E_{max} = 10^6$ MeV, $n_H = 300$ cm$^{-3}$,
$B = 0.29 G$;  short dashed curve -- case B:
same as Case A but with $E_{max} = 10^5$ MeV; 
dot-dashed curve -- case C: same as Case A but with $\alpha_p = 2.4$; 
Case D (same as case A but with $n_H = 3,000$ cm$^{-3}$ and $B = 0.45$ mG)
produces the same curve as case A, and is therefore also 
represented by the long dashed curve.
}
\end{figure}

\clearpage\newpage
\begin{figure}
\figurenum{5}
{\epsscale{0.9} \plotone{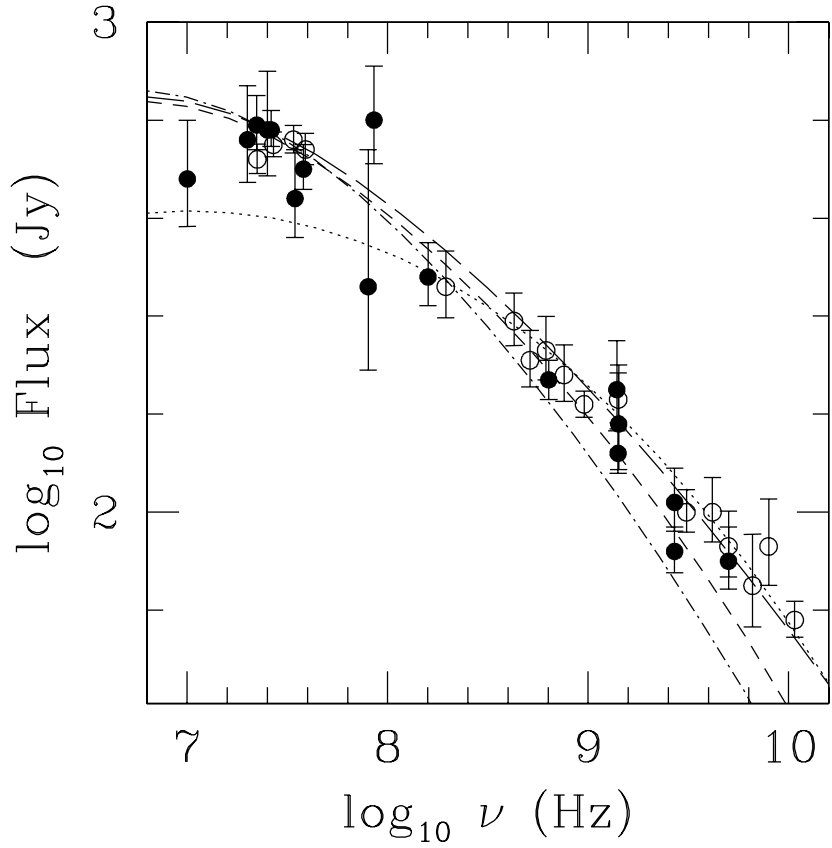} } 
\figcaption{Radio synchrotron emission produced by secondary leptons
associated with the four cases presented in Figure 4.  Long dashed 
curve: case A; short dashed curve: case B; dot-dashed curve: case C; 
dotted curve: case D.  Radio data for IC443 taken from Erickson
\& Mahoney 1985.
}
\end{figure}

\clearpage\newpage
\begin{figure}
\figurenum{6}
{\epsscale{0.9} \plotone{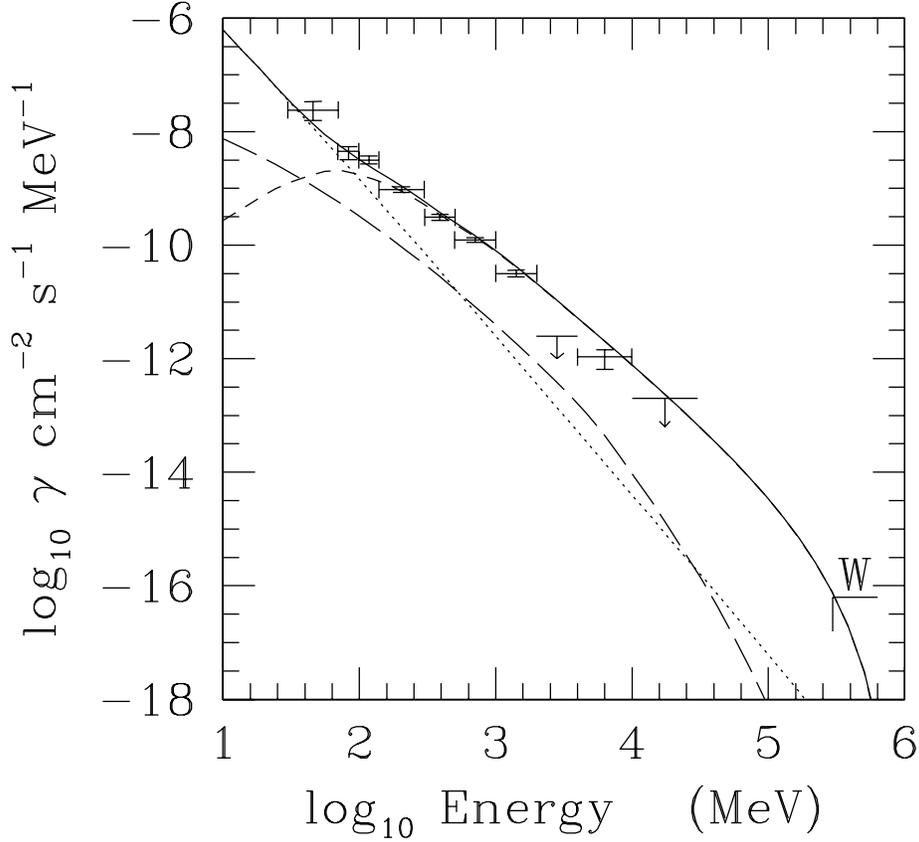} }
\figcaption{The $\gamma$-ray emissivity from our pion-decay model
with $\alpha_p = 2$, $E_{max} = 10^6$ MeV, $n_H = 300$ cm$^{-3}$
and $B = 0.29$ mG and as assumed primary electron population with
spectral index of $\alpha_e = 2.8$.  The remnant's
age is assumed to be $30,000$ years in order to determine the secondary
leptons' (non steady-state) distribution. Short-dashed line: photons
produced via the decay of neutral pions; long dashed curve: bremsstrahlung
emission from the secondary leptons; dotted line: bremsstrahlung emission
from the primary electron population.
}
\end{figure}

\clearpage\newpage
\begin{figure}
\figurenum{7}
{\epsscale{0.9} \plotone{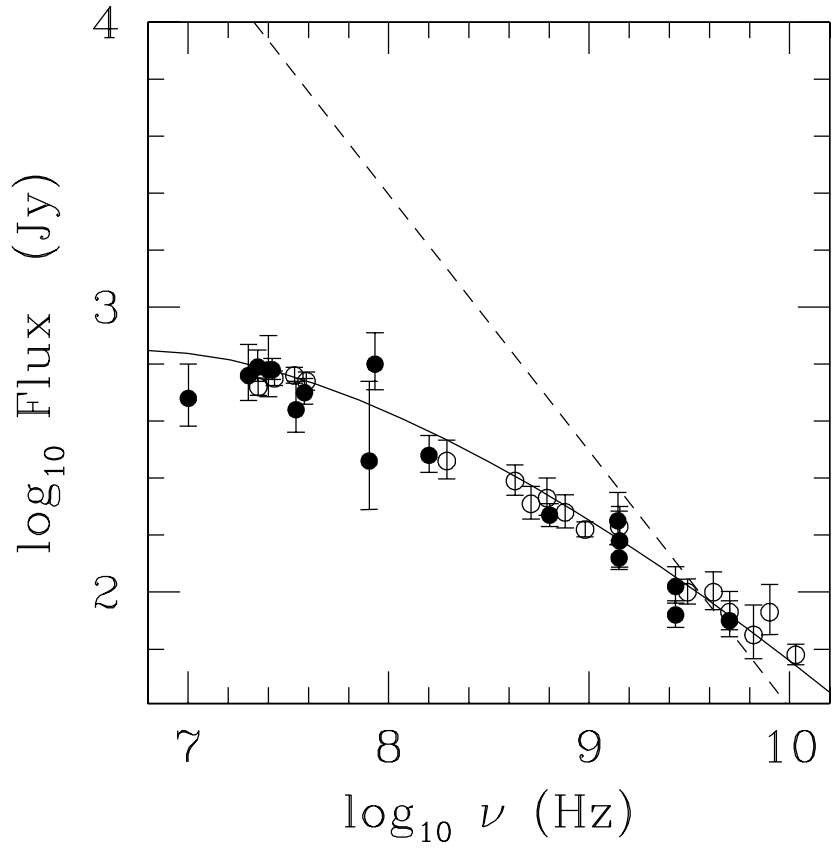} }
\figcaption{The synchrotron emission produced by the 
primary electrons (dashed curve) and secondary leptons
(solid curve) associated with the curves presented in 
Figure 6 for a field strength of $B = 0.29$ mG.  
}
\end{figure}

\clearpage\newpage
\begin{figure}
\figurenum{8}
{\epsscale{0.9} \plotone{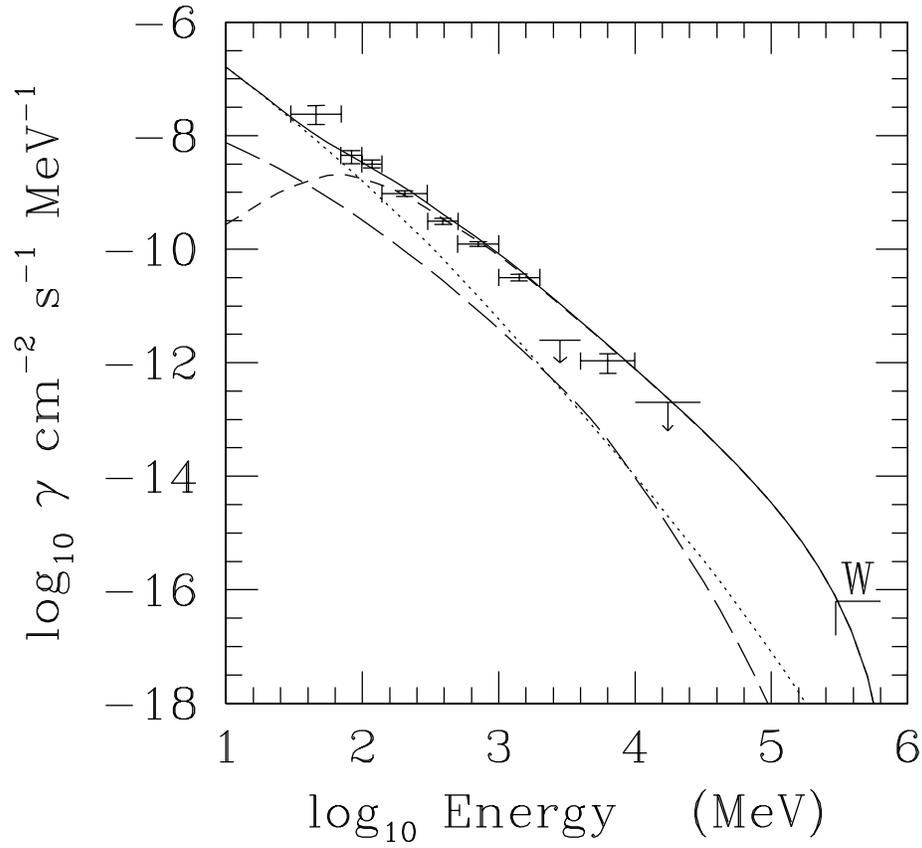} }
\figcaption{Same as Figure 6, but for a steady-state
population of primary electrons injected into the
dense core environment with a spectral index of $\alpha_e
= 2.8$ at the same energy rate as the protons.  
}
\end{figure}

\clearpage\newpage
\begin{figure}
\figurenum{9}
{\epsscale{0.9} \plotone{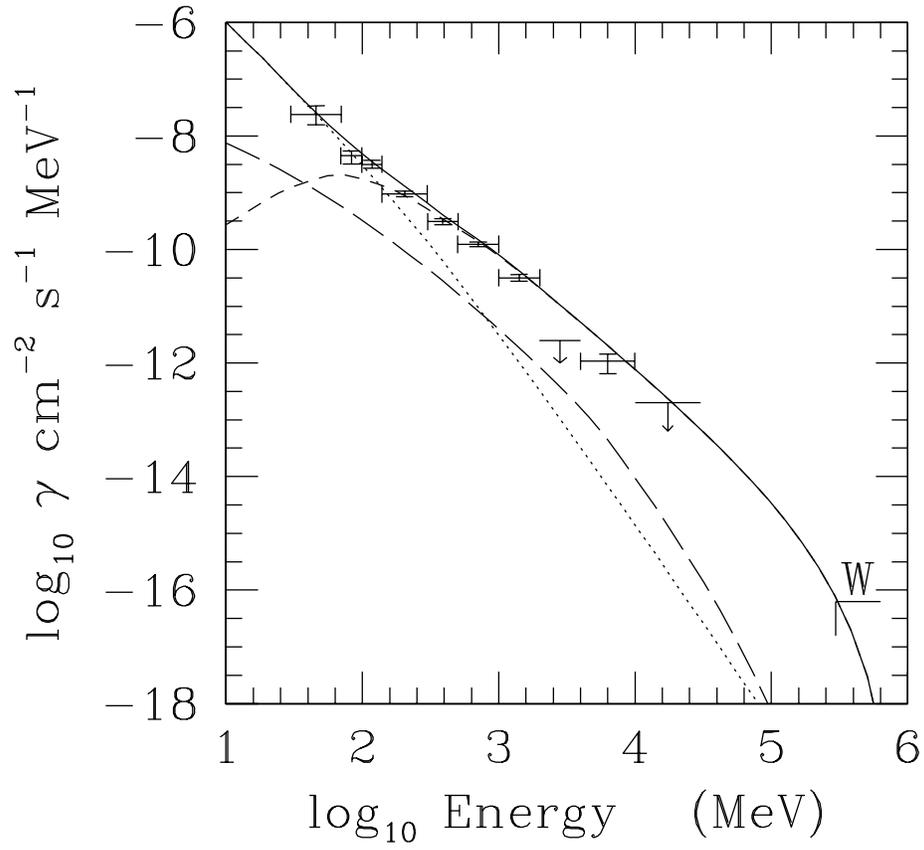} }
\figcaption{Same as Figure 8, but with $\alpha_e = 3.4$
above 10 MeV.
}
\end{figure}

\clearpage\newpage
\begin{figure}
\figurenum{10}
{\epsscale{0.9} \plotone{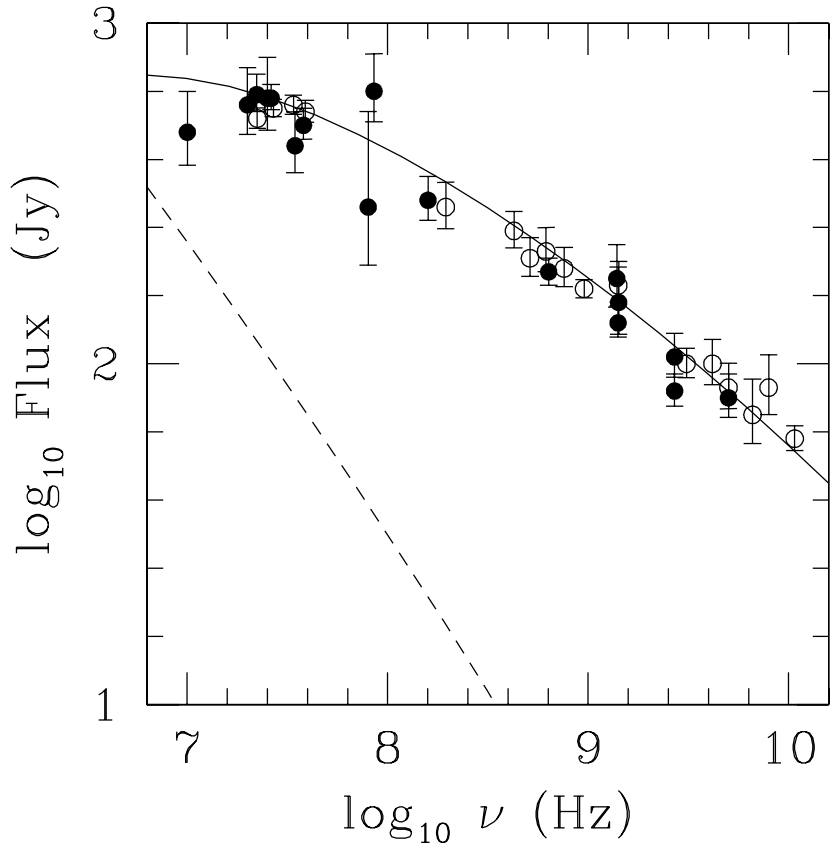} }
\figcaption{The synchrotron emission produced by the primary 
electrons (dashed curve) and secondary leptons (solid curves) 
associated with the $\gamma$-ray fits presented in Figure 9.
The primary electrons are assumed to populate the
same region as the secondary leptons, as a result of their
diffusion (after cooling) from the dense cores.
}
\end{figure}

\clearpage\newpage
\begin{figure}
\figurenum{11}
{\epsscale{0.9} \plotone{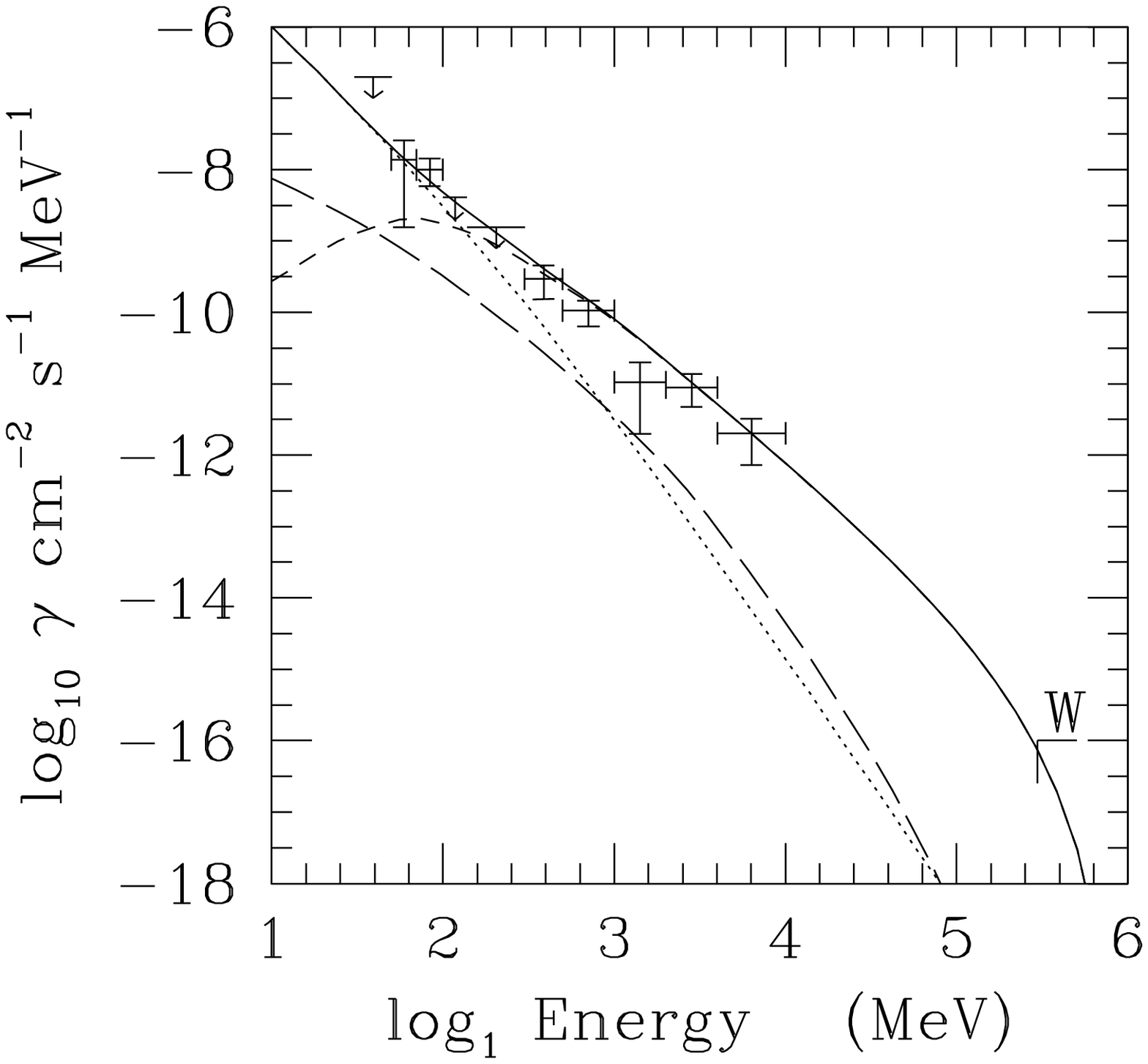} }
\figcaption{The $\gamma$-ray spectrum produced by
the two-zone model applied to W44, where
$\alpha_p = 2$ and $B = 0.43$ mG.  Short dashed curve:
neutral-pion decay emissivity; long-dashed curve: 
bremsstrahlung emissivity produced by secondary 
leptons; dotted curve: bremsstrahlung produced by 
primary electrons; solid line: total $\gamma$-ray 
emissivity. 
Data are taken from Merck et al. (1996) and Buckley
et al. (1998).}
\end{figure}

\clearpage\newpage
\begin{figure}
\figurenum{12}
{\epsscale{0.9} \plotone{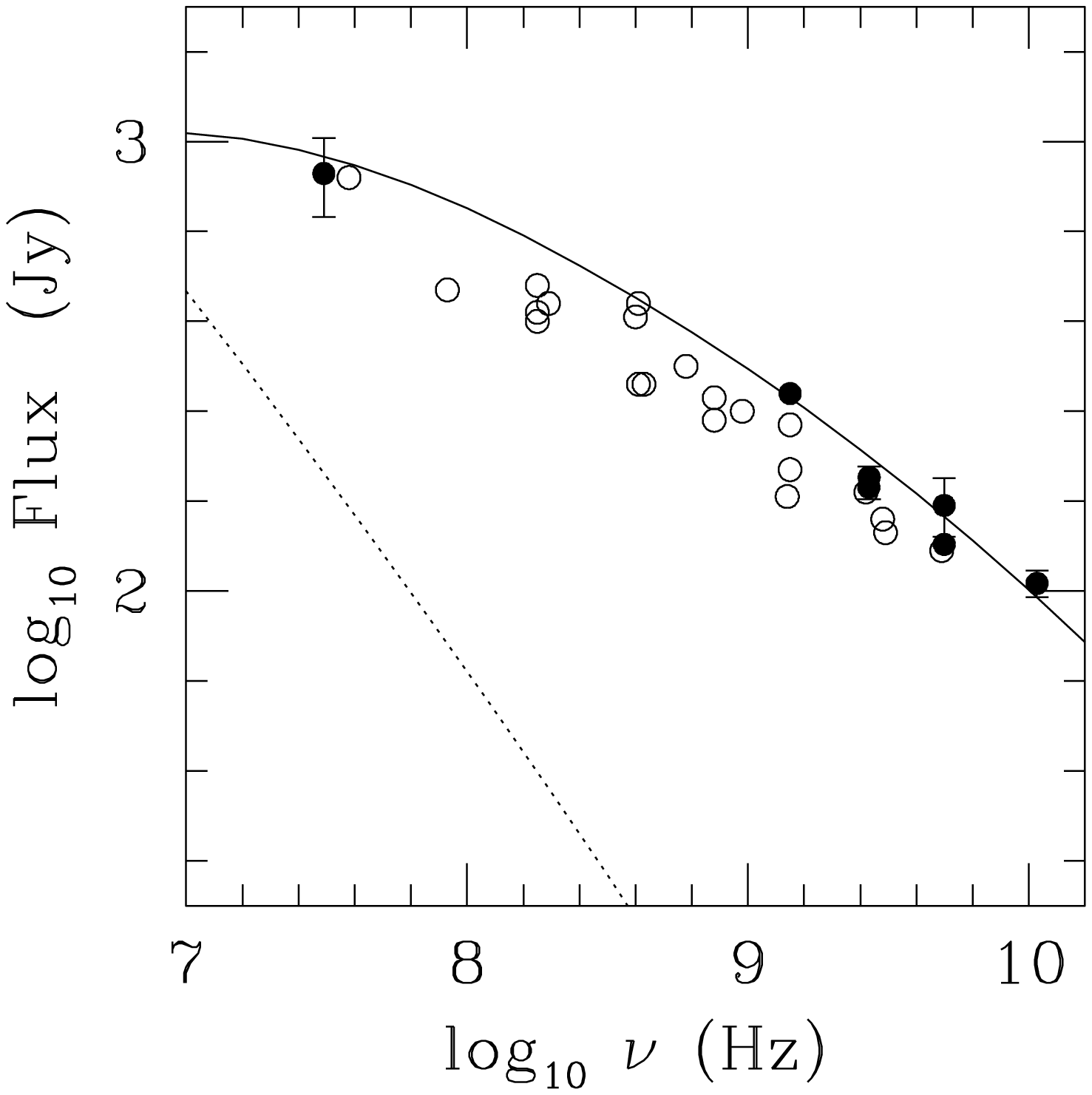} }
\figcaption{The radio spectrum produced by the two-zone
model applied to W44, where $\alpha_p = 2$
and $B = 0.43$ mG.  Dashed curve: synchrotron emission produced 
by the primary electrons that have diffused out into the shell; 
solid curve: synchrotron emission produced by the secondary leptons.
Radio data taken from Kassim (1989). Solid circles represent data
for which an error was quoted.  Open circles represent data for which
not data was quoted.
}
\end{figure}

\clearpage\newpage
\begin{figure}
\figurenum{13}
{\epsscale{0.9} \plotone{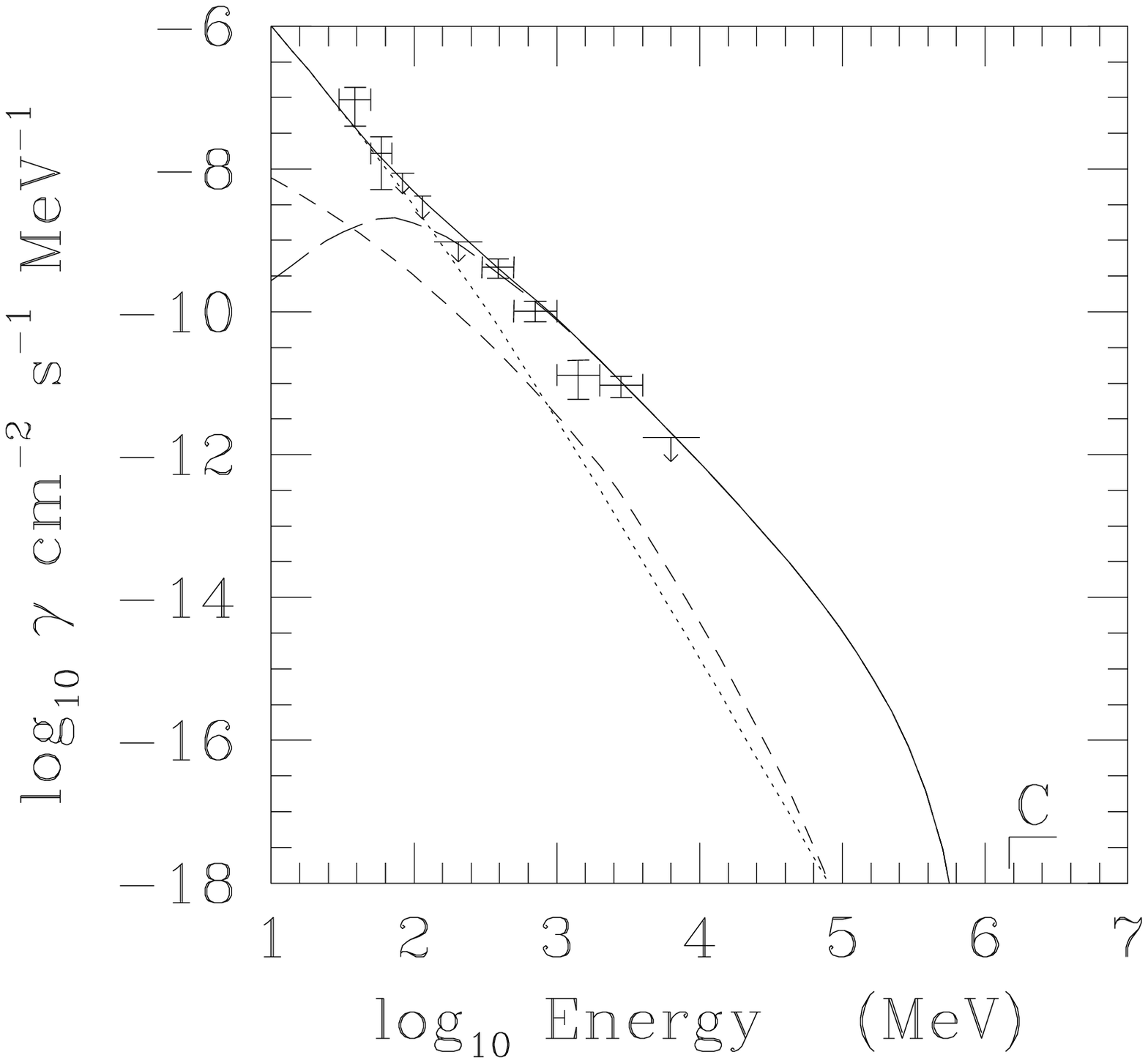} }
\figcaption{The $\gamma$-ray spectrum produced by
the two-zone model applied to W28, where
$\alpha_p = 2$ and $B = 0.43$ mG.  Short dashed curve:
neutral-pion decay emissivity; long-dashed curve: 
bremsstrahlung emissivity produced by secondary 
leptons; dotted curve: bremsstrahlung produced by 
primary electrons; solid line: total $\gamma$-ray 
emissivity. 
Data are taken from Merck et al. (1996) and Rowell et al.
(2000).}
\end{figure}

\clearpage\newpage
\begin{figure}
\figurenum{14}
{\epsscale{0.9} \plotone{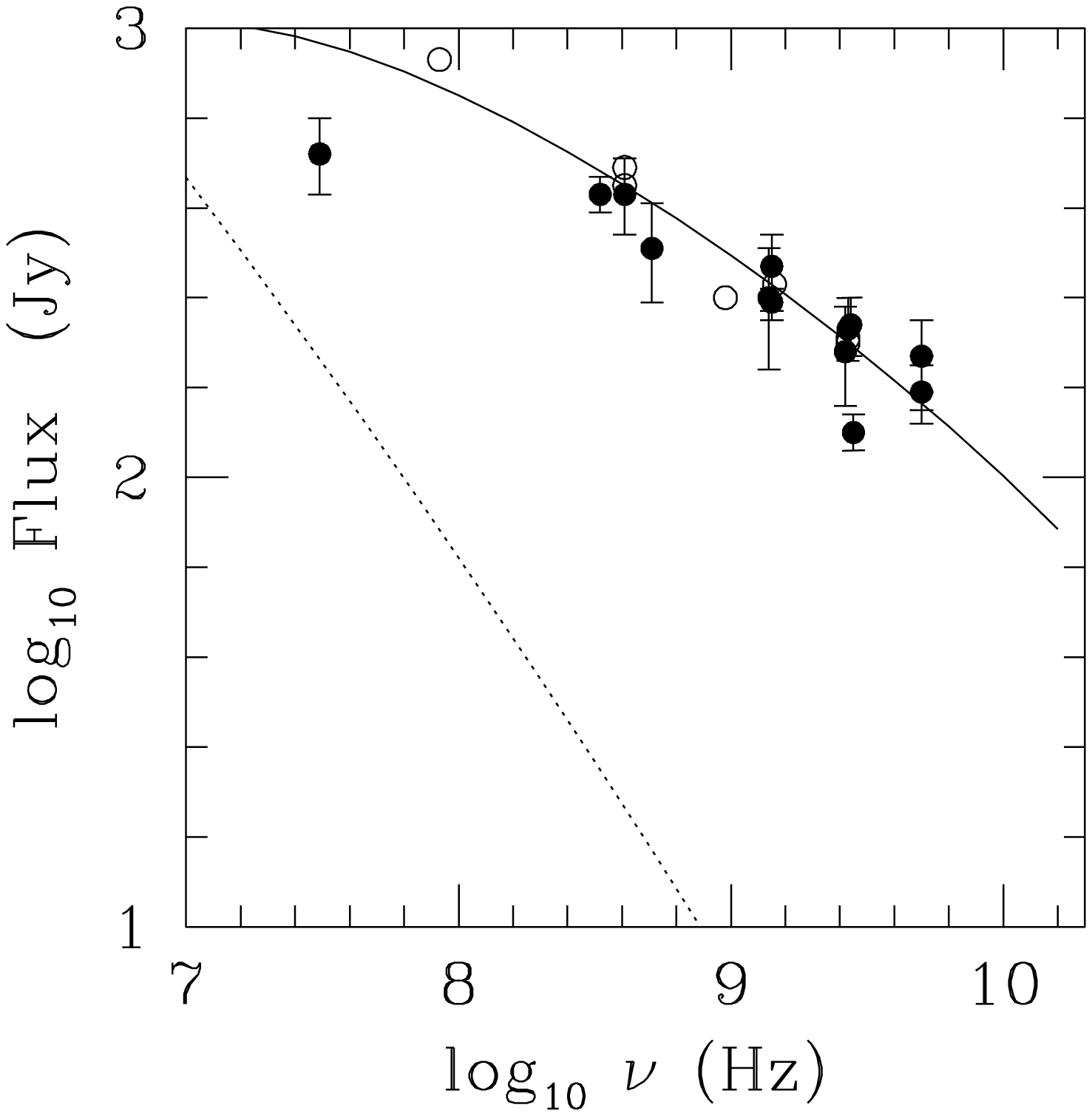} }
\figcaption{The radio spectrum produced by the two-zone
model applied to W28, where $\alpha_p = 2$
and $B = 0.43$ mG.  Dashed curve: synchrotron emission produced 
by the primary electrons that have diffused out into the shell; 
solid curve: synchrotron emission produced by the secondary leptons.
Radio data taken from Kassim (1989).  Solid circles represent
data for which an error was quoted.  Open circles represent
data for which no error was quoted.
}
\end{figure}

\clearpage\newpage
\begin{figure}
\figurenum{15}
{\epsscale{0.9} \plotone{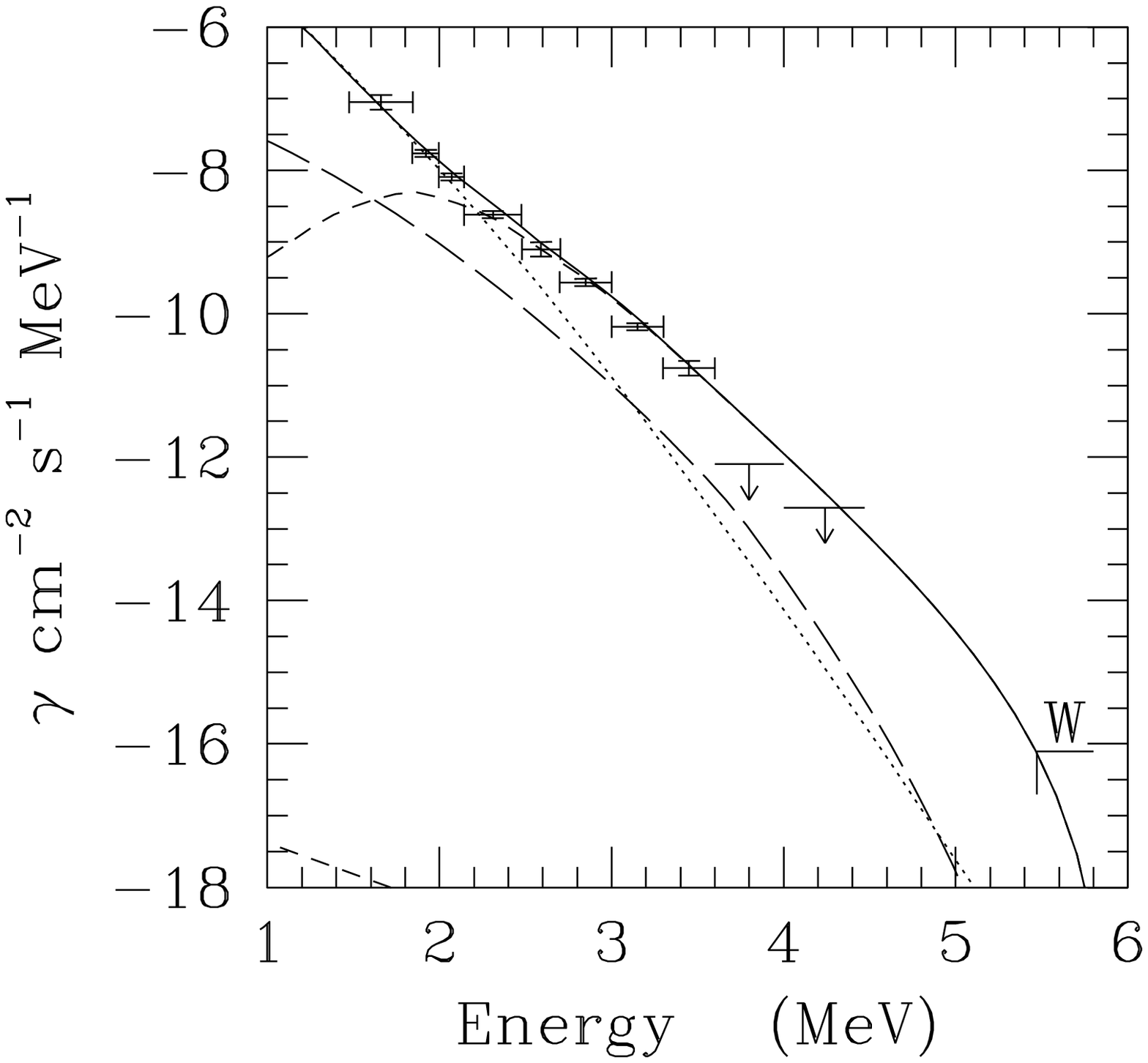} }
\figcaption{The $\gamma$-ray spectrum produced by
the two-zone model applied to $\gamma$-Cygni, where
$\alpha_p = 2.2$ and $B = 0.28$ mG.  Short dashed curve:
neutral-pion decay emissivity; long-dashed curve: 
bremsstrahlung emissivity produced by secondary 
leptons; dotted curve: bremsstrahlung produced by 
primary electrons; solid line: total $\gamma$-ray 
emissivity. 
Data are taken from Merck et al. (1996) and Buckley et al.
1998.}
\end{figure}

\clearpage\newpage
\begin{figure}
\figurenum{16}
{\epsscale{0.9} \plotone{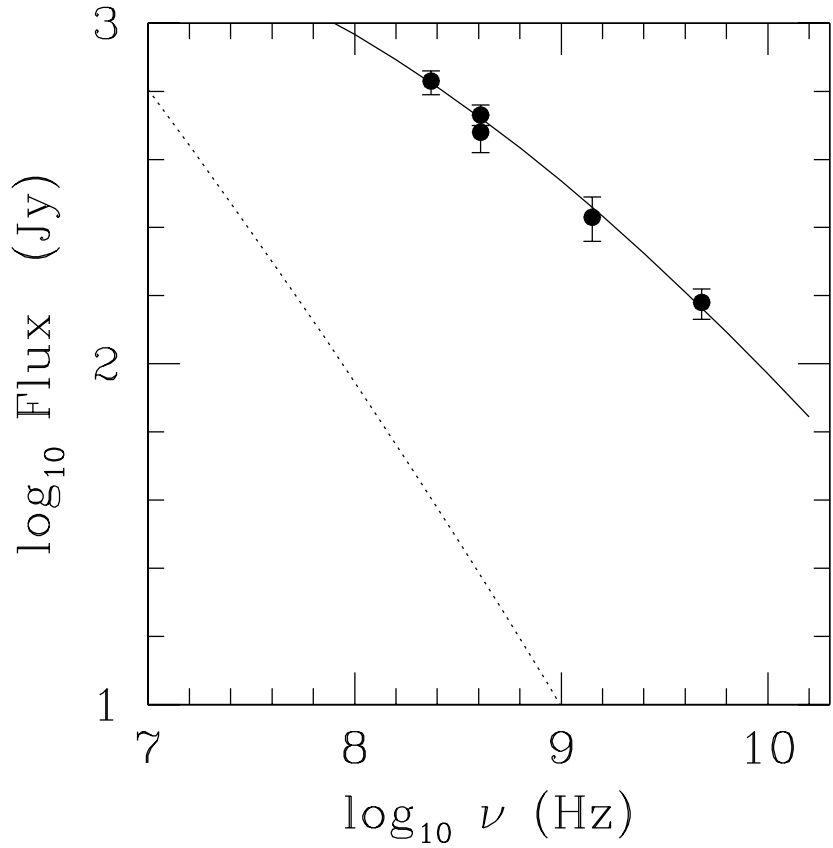} }
\figcaption{The radio spectrum produced by the two-zone
model applied to $\gamma$-Cygni, where $\alpha_p = 2.2$
and $B = 0.28$ mG.  Dashed curve: synchrotron emission produced 
by the primary electrons that have diffused out into the shell; 
solid curve: synchrotron emission produced by the secondary leptons.
Radio data taken from Zhang et al. (1997).
}
\end{figure}

\end{document}